\documentstyle[11pt,newpasp,twoside,epsf,psfig]{article}
\def\farcm{\hbox{$.\mkern-4mu^\prime$}}
\def\farcs{\hbox{$.\!\!^{\prime\prime}$}}

\begin{document}
\title{Ionized Gas in Spheroids: The SAURON Survey} \author{D.\
 Krajnovi\'c\altaffilmark{1}$^{,8}$, R.\ Bacon$^2$, M.\ Bureau$^1$, C.M.\
 Carollo$^3$, Y.\ Copin$^1$, R.L.\ Davies$^4$, E.\ Emsellem$^2$, H.\
 Kuntschner$^4$, R.\ McDermid$^4$, B.W.\ Miller$^5$, G.\ Monnet$^6$, 
 R.F.\ Peletier$^7$, E.K.\ Verolme$^1$, P.T.\ de Zeeuw$^1$} 




\altaffiltext{1}{Sterrewacht Leiden, Leiden University, 
            $^2$CRAL--Observatoire de Lyon, Saint--Genis--Laval, 
            $^3$Columbia University, New York,
            $^4$Physics Department, University of Durham, 
            $^5$Gemini International Observatory, La Serena,
            $^6$European Southern Observatory, Garching,
            $^7$School of Physics and Astronomy, University of Nottingham,
            $^8$Institute Rudjer Boskovic, Zagreb}

\begin{abstract}
Early results are reported from the SAURON survey of the kinematics
and stellar populations of nearby cluster and field E, S0, and Sa
galaxies. We present maps of the H$\beta$ and [OIII] emission-line
distribution and kinematics for NGC 5813 and NGC 7742.
\end{abstract}

\section{Introduction}

Studies with HST have revealed a variety of structures in the nuclei
of nearby elliptical galaxies: stellar and/or gaseous disks, massive
central black holes, unresolved nuclear spikes, and kinematically
decoupled structures. Progress towards a better understanding of the
properties of nearby galaxies requires a systematic investigation of
the large-scale (ground-based) kinematics and line-strengths of a
representative sample. Traditionally such studies relied on long-slit
spectroscopy along at most a few position angles, but this is
insufficient to unravel the rich internal kinematics of spheroids. For
this reason we built the panoramic integral-field spectrograph SAURON
for the William Herschel Telescope.\looseness=-2

The design of SAURON is similar to that of the prototype
integral-field spectrograph TIGER and its successor OASIS built for
the CFHT (Bacon et al.\ 1995, 2000).  The field of view is $33\arcsec
\times 41\arcsec$ with spatial sampling of $0\farcs94 \times
0\farcs94$.  SAURON provides 1577 spectra in the wavelength range from
4810 to 5350\AA. Of these, 146 are sky spectra $1\farcm9$ away from
the main field. A complete description of the design and construction
of SAURON, and the corresponding data reduction software is given in
Bacon et al.\ (2001).

We are carrying out a systematic study of nearby cluster and field E,
S0 and Sa galaxies. The aim is to determine the intrinsic shapes of
the galaxies, their orbital structure, the mass-to-light ratio as a
function of radius, the age and metallicity of the stellar components,
the dynamical role of decoupled cores and nuclear black holes, the
relation between the stellar (and gaseous) kinematics, and the star
formation history. Here we illustrate the variety of gas morphologies
and kinematics displayed by nearby spheroids by presenting SAURON maps
of the H$\beta$ and [OIII] emission-line gas for one elliptical and
one spiral galaxy of our sample, NGC 5813 and NGC 7742,
respectively.\looseness=-2

\begin{figure}
\centerline{\psfig{file=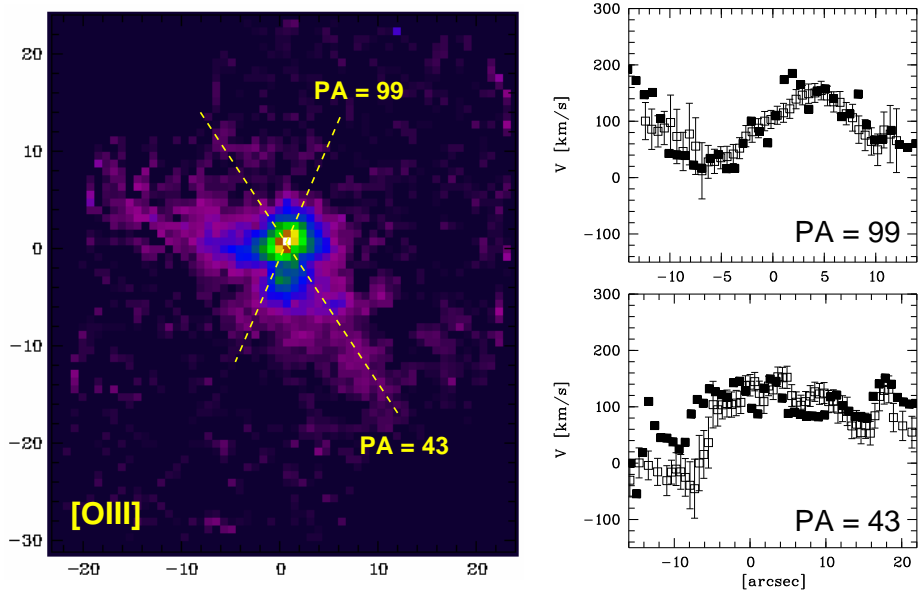,angle=-90.,width=12.2cm,clip=}}
\caption{Left: intensity of the [OIII] emission in the galaxy NGC
5813, as measured with SAURON. The map is based on two pointings of
$4\times 1800$~s each, sampled with $0\farcs8\times0\farcs8$
pixels. Right: comparison of the SAURON measurements with the
H$\alpha$+N[II] long-slit spectroscopy of Caon et al.\ (2000) taken
along the dashed lines in the [OIII] map. }
\end{figure}

\section{The giant elliptical galaxy NGC 5813}

NGC 5813 is an E1-2 galaxy located in the Virgo-Libra Cloud (Tully
1988) with absolute magnitude $M_B=-20.99$. It contains a weak
unresolved radio continuum source (Birkinshaw \& Davies 1985;
Krajnovic \& Jaffe, in prep.). It has typical LINER emission-line
ratios, but no X-ray emission (Ho, Filippenko \& Sargent 1997). HST
imaging shows dust filaments inside the kinematically decoupled core
(Carollo et al.\ 1997). Ground-based H$\alpha$+[NII] narrow-band
imaging reveals an elongated ionized gas distribution (Caon et al.\
2000).\looseness=-2

We observed NGC 5813 on March 29 and 30 and April 4, 2000. The seeing
was 1\arcsec\ on the first two nights and $2\farcs5$ on the
last. Figure 1 shows the total intensity of the [OIII] emission,
together with a comparison of the SAURON data and the Caon et al.\
(2000) long-slit data along two position angles.\footnote{Colour
versions of Figures 1 and 2 can be found in de Zeeuw et al.\ (2001).}
The map is a mosaic obtained by merging two SAURON pointings which
overlap on the nucleus. The morphology of the [OIII] emission exhibits
complex filamentary structure, most likely not yet in equilibrium with
the potential of the central region, despite the relatively short
crossing time of $\approx$10$^7$~yr.  This is confirmed by the
disorganized SAURON velocity field (see also de Zeeuw et al.\ 2001).

\vfill\eject

\begin{figure}
\centerline{\psfig{file=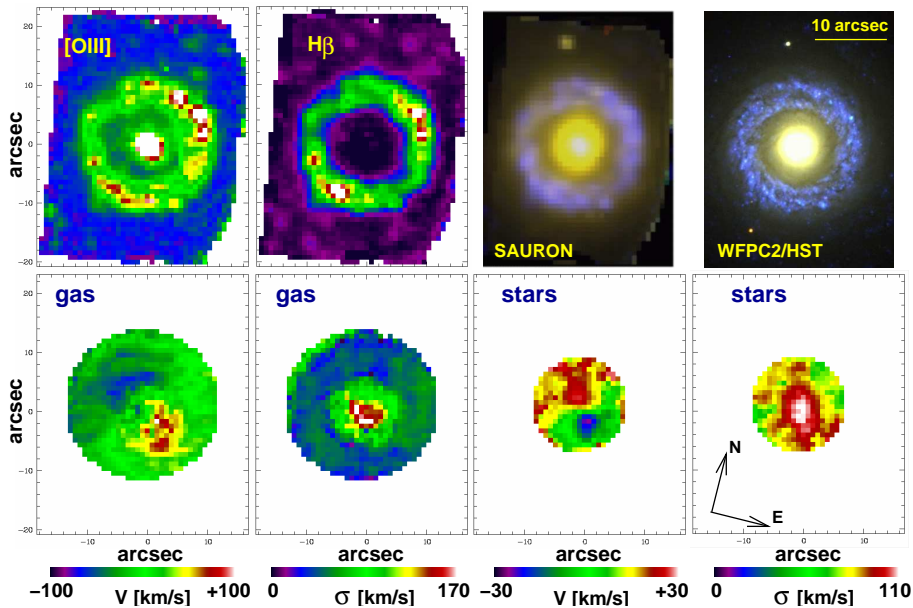,width=12.0cm}}
\caption{SAURON maps of NGC 7742, based on one pointing, exposed for
$3\times1800 + 1\times 900$~s. The top panels show the emission-line
intensity distributions of O[III] and H$\beta$, followed by a
colour-coded reconstructed image composed of [OIII] (blue), blue
continuum (green), red continuum (red) derived from the SAURON data,
and a similar colour-coded image composed of HST/WFPC2 exposures with
the F336W (blue), F555W (green) and F814 (red) filters.  The bottom
row shows (from left to right) the derived gas velocity and velocity
dispersion fields, and the stellar velocity and velocity dispersion
fields. }
\end{figure}

\section{Counter-rotating gas in the spiral galaxy NGC 7742}

NGC 7742 is a nearly face-on spiral galaxy classified as S(r)b in the
RC3 ($M_B=-20.99$), and it is an example of the latest spirals
included in our sample. The galaxy contains an inner stellar ring with
bright HII regions, flocculent spiral arms, a significant amount of
neutral hydrogen, molecular gas, and dust (de Vaucouleurs \& Buta
1980; Roberts et al.\ 1991; Wozniak et al.\ 1995).  The nucleus is
classified as a transition LINER/HII object (Ho et al.\ 1997).

Figure 2 displays SAURON maps based on a single pointing taken on
October 13, 1999, with a seeing of $\approx 2''$. Most of the H$\beta$
and [OIII] emission is confined to the ring of star formation
surrounding the bulge. H$\beta$ is dominant everywhere in the ring,
and the ratio [OIII]/H$\beta$ ranges from 0.06 to 0.14. In the center
the ratio is larger than 20.  Figure 2 also shows a colour-coded
SAURON reconstructed image and a similar image based on WFPC2
exposures. The SAURON map does not have the spatial resolution of HST,
but it does demonstrate that our analysis technique is capable of
providing accurate emission-line maps.\looseness=-2

The bottom panels of Figure 2 display the mean velocity and velocity
dispersion fields of the gas and the stars in NGC 7742. The galaxy is
observed close to face-on, and the amplitude of the velocities is
accordingly modest. The zero-velocity directions are well-defined, and
their position angles are consistent with each other (PA=$42^\circ\pm
12^\circ$ for the stars and PA=$35^\circ\pm5^\circ$ for the
gas). However, the gas and stars rotate in opposite sense. This
suggests a merger or an accretion event in the history of the galaxy.

\section{Conclusion}

We built SAURON to efficiently measure the stellar and gaseous
kinematics and linestrength distributions of early-type galaxies.  The
two examples shown here illustrate SAURON's ability to measure the
two-dimensional emission-line distributions. The results presented in
de Zeeuw et al.\ (2001) show that early-type galaxies display a
diversity of line-strength distributions and kinematical structures
much richer than previously assumed.  Integral-field spectroscopy is a
mature technique, superior to long-slit studies because of the
fundamental advantage of complete spatial coverage.

The detailed SAURON maps, together with high spatial resolution data
from OASIS and HST, will be used for modeling of the dynamics by means
of Schwarzschild's (1979) orbit superposition method.  When combined
with the constraints on the stellar populations derived from the
line-strength distributions this will provide much needed insight into
the fundamental properties of early-type galaxies and bulges.

\medskip
It is a pleasure to thank the ING staff for enthusiastic and competent
support on La Palma.

\end{document}